\documentclass[aps,prd,superscriptaddress,preprint,tightenlines,nofootinbib]{revtex4}

\usepackage{graphicx}
\usepackage{dcolumn}
\usepackage{bm}

\newcommand{\text}{\mathrm}

\begin{document}

\preprint{CLNS 09/2054}
\preprint{CLEO 09-07}

\title{\boldmath Measurements of $D$ Meson Decays to Two Pseudoscalar Mesons}

\author{H.~Mendez}
\affiliation{University of Puerto Rico, Mayaguez, Puerto Rico 00681}
\author{J.~Y.~Ge}
\author{D.~H.~Miller}
\author{I.~P.~J.~Shipsey}
\author{B.~Xin}
\affiliation{Purdue University, West Lafayette, Indiana 47907, USA}
\author{G.~S.~Adams}
\author{D.~Hu}
\author{B.~Moziak}
\author{J.~Napolitano}
\affiliation{Rensselaer Polytechnic Institute, Troy, New York 12180, USA}
\author{K.~M.~Ecklund}
\affiliation{Rice University, Houston, Texas 77005, USA}
\author{Q.~He}
\author{J.~Insler}
\author{H.~Muramatsu}
\author{C.~S.~Park}
\author{E.~H.~Thorndike}
\author{F.~Yang}
\affiliation{University of Rochester, Rochester, New York 14627, USA}
\author{M.~Artuso}
\author{S.~Blusk}
\author{S.~Khalil}
\author{R.~Mountain}
\author{K.~Randrianarivony}
\author{T.~Skwarnicki}
\author{S.~Stone}
\author{J.~C.~Wang}
\author{L.~M.~Zhang}
\affiliation{Syracuse University, Syracuse, New York 13244, USA}
\author{G.~Bonvicini}
\author{D.~Cinabro}
\author{A.~Lincoln}
\author{M.~J.~Smith}
\author{P.~Zhou}
\author{J.~Zhu}
\affiliation{Wayne State University, Detroit, Michigan 48202, USA}
\author{P.~Naik}
\author{J.~Rademacker}
\affiliation{University of Bristol, Bristol BS8 1TL, UK}
\author{D.~M.~Asner}
\author{K.~W.~Edwards}
\author{J.~Reed}
\author{A.~N.~Robichaud}
\author{G.~Tatishvili}
\author{E.~J.~White}
\affiliation{Carleton University, Ottawa, Ontario, Canada K1S 5B6}
\author{R.~A.~Briere}
\author{H.~Vogel}
\affiliation{Carnegie Mellon University, Pittsburgh, Pennsylvania 15213, USA}
\author{P.~U.~E.~Onyisi}
\author{J.~L.~Rosner}
\affiliation{University of Chicago, Chicago, Illinois 60637, USA}
\author{J.~P.~Alexander}
\author{D.~G.~Cassel}
\author{R.~Ehrlich}
\author{L.~Fields}
\author{L.~Gibbons}
\author{S.~W.~Gray}
\author{D.~L.~Hartill}
\author{B.~K.~Heltsley}
\author{J.~M.~Hunt}
\author{J.~Kandaswamy}
\author{D.~L.~Kreinick}
\author{V.~E.~Kuznetsov}
\author{J.~Ledoux}
\author{H.~Mahlke-Kr\"uger}
\author{J.~R.~Patterson}
\author{D.~Peterson}
\author{D.~Riley}
\author{A.~Ryd}
\author{A.~J.~Sadoff}
\author{X.~Shi}
\author{S.~Stroiney}
\author{W.~M.~Sun}
\author{T.~Wilksen}
\affiliation{Cornell University, Ithaca, New York 14853, USA}
\author{J.~Yelton}
\affiliation{University of Florida, Gainesville, Florida 32611, USA}
\author{P.~Rubin}
\affiliation{George Mason University, Fairfax, Virginia 22030, USA}
\author{N.~Lowrey}
\author{S.~Mehrabyan}
\author{M.~Selen}
\author{J.~Wiss}
\affiliation{University of Illinois, Urbana-Champaign, Illinois 61801, USA}
\author{M.~Kornicer}
\author{R.~E.~Mitchell}
\author{M.~R.~Shepherd}
\author{C.~M.~Tarbert}
\affiliation{Indiana University, Bloomington, Indiana 47405, USA }
\author{D.~Besson}
\affiliation{University of Kansas, Lawrence, Kansas 66045, USA}
\author{T.~K.~Pedlar}
\author{J.~Xavier}
\affiliation{Luther College, Decorah, Iowa 52101, USA}
\author{D.~Cronin-Hennessy}
\author{K.~Y.~Gao}
\author{J.~Hietala}
\author{T.~Klein}
\author{R.~Poling}
\author{P.~Zweber}
\affiliation{University of Minnesota, Minneapolis, Minnesota 55455, USA}
\author{S.~Dobbs}
\author{Z.~Metreveli}
\author{K.~K.~Seth}
\author{B.~J.~Y.~Tan}
\author{A.~Tomaradze}
\affiliation{Northwestern University, Evanston, Illinois 60208, USA}
\author{S.~Brisbane}
\author{J.~Libby}
\author{L.~Martin}
\author{A.~Powell}
\author{C.~Thomas}
\author{G.~Wilkinson}
\affiliation{University of Oxford, Oxford OX1 3RH, UK}
\collaboration{CLEO Collaboration}
\noaffiliation

\date{June 17, 2009}

\begin{abstract} 
Using data collected on the $\psi(3770)$ resonance and near the
$D^{\ast \pm}_s D^{\mp}_s$ peak production energy by the CLEO-c detector, we
study the decays of the possible $D \rightarrow PP$ modes and report
measurements of or upper limits on all branching fractions for
Cabibbo-favored, singly-Cabibbo-suppressed, and
doubly-Cabibbo-suppressed $D \rightarrow PP$ decays except modes
involving $K^0_L$ (and except $D^0 \rightarrow K^+ \pi^-$). We
normalize with respect to the Cabibbo-favored $D$ modes, $D^0 \rightarrow K^-
\pi^+$, $D^+ \rightarrow K^- \pi^+ \pi^+$, and $D^+_s \rightarrow
K^+K^{0}_{S}$.
\end{abstract}

\maketitle

\section{\boldmath Introduction}

There are many possible exclusive decays of charmed $D$ mesons to a pair of
mesons from the lowest-lying pseudoscalar meson nonet. The decay can
be to any pair of $K^+$, $K^-$, $\pi^+$, $\pi^-$, $\eta$, $\eta'$,
$\pi^0$, $K^0$, or $\bar{K^0}$, with total charge 0 or $\pm
1$. Measurements of the complete set of
decays can be used to test flavor topology and SU(3) predictions and
to specify strong phases of decay amplitudes through triangle
relations~\cite{RosnerPaper}. Moreover, many $CP$ asymmetries
(expected to be less than ${\cal O}(10^{-3})$ in the Standard Model)
can be studied. The detectable neutral kaons
are $K^0_S$ and $K^0_L$, not $K^0$ and $\bar{K^0}$, so the observable
decays are $X K^0_S$ and $X K^0_L$. In this study, we consider only
$K^0_S$, not $K^0_L$, and report all branching fractions for
Cabibbo-favored, singly-Cabibbo-suppressed, and
doubly-Cabibbo-suppressed $D \rightarrow PP$ decays except modes
involving $K^0_L$ and except the doubly-Cabibbo-suppressed decay $D^0
\rightarrow K^+ \pi^-$. We normalize with respect to the Cabibbo-favored $D$
modes, $D^0 \rightarrow K^- \pi^+$~\cite{CLEO:sys}, $D^+ \rightarrow
K^- \pi^+ \pi^+$~\cite{CLEO:sys}, and $D^+_s \rightarrow
K^+K^{0}_{S}$~\cite{Peterpaper}. (More precisely, we normalize the
$D^0 \rightarrow PP$ decays with respect to the sum of the Cabibbo-favored
mode $D^0 \rightarrow K^- \pi^+$ and the doubly-Cabibbo-suppressed
mode $D^0 \rightarrow K^+ \pi^-$. The latter is 0.4\% of the former.)

\section{\boldmath The detector}

Data for this analysis were taken at the Cornell Electron Storage 
Ring (CESR) using the CLEO-c general-purpose solenoidal detector,
which is described in detail
elsewhere~\cite{Briere:2001rn,Kubota:1991ww,cleoiiidr,cleorich}. 
The charged particle tracking system covers a solid angle of 93\% of
$4 \pi$ and consists of a small-radius, six-layer, low-mass, stereo
wire drift chamber, concentric with, and surrounded by, a 47-layer
cylindrical central drift chamber. The chambers operate in a 1.0 T
magnetic field. The root-mean-square (rms) momentum resolution
achieved with the tracking system is approximately 0.6\% at
$p=1$~GeV/$c$ for tracks that traverse all layers of the drift
chamber.
Photons are detected in an electromagnetic calorimeter consisting of
7800 cesium iodide crystals and covering 95\% of $4 \pi$, which
achieves a photon energy resolution of 2.2\% at $E_\gamma=1$~GeV and
6\% at 100~MeV.
We utilize two particle identification (PID) devices to separate
charged kaons from pions: the central drift chamber, which provides
measurements of ionization energy loss ($dE/dx$), and, surrounding
this drift chamber, a cylindrical ring-imaging Cherenkov (RICH)
detector, whose active solid angle is 80\% of $4 \pi$. The combined
PID system has a pion or kaon efficiency $>85\%$ and a probability of
pions faking kaons (or vice versa) $<5\%$~\cite{CLEO:sys}.
The response of the CLEO-c detector is studied with a detailed
GEANT-based~\cite{geant} Monte Carlo (MC) simulation, with initial
particle trajectories generated by EvtGen~\cite{evtgen} and
final state radiation produced by PHOTOS~\cite{photos}. Simulated
events are reconstructed and selected for analysis with the
reconstruction programs and selection criteria used for data.

\section{\boldmath The Data Sample}

For $D^0$ and $D^+$ meson decays, we utilize a total integrated
luminosity of 818 $\mathrm{pb}^{-1}$ of $e^+e^-$ data collected at
center-of-mass (CM) energies near $E_{\text{cm}}=3774$~MeV. The data
sample contains about $2.4 \times 10^{6}$ $D^{+} D^{-}$ events (events
of interest), three million $D^{0} \bar{D^{0}}$ events (events of
interest), fifteen million $e^{+} e^{-} \rightarrow u \bar{u},\ d
\bar{d},$ or $s \bar{s}$ continuum events, three million 
$e^{+} e^{-} \rightarrow \tau^{+} \tau^{-}$ events, and three million
$e^{+} e^{-} \rightarrow \gamma \psi^{\prime}$ radiative return events 
(sources of background), as well as Bhabha events, $\mu$-pair events,
and $\gamma \gamma$ events (useful for luminosity determination and
resolution studies). For the $D_s^+$ meson decays, we use a data
sample of $e^+e^- \rightarrow D^{\ast \pm}_s D^{\mp}_s$ events
collected at the CM energy 4170~MeV, near $D^{\ast \pm}_s D^{\mp}_s$
peak production of $\sim$1 nb~\cite{Poling:2006da}. The data sample
consists of an integrated luminosity of 586 $\mathrm{pb}^{-1}$
containing about $5.4 \times 10^5$ $D^{\ast \pm}_s D^{\mp}_s$
pairs. Other charm production totals $\sim$7 nb~\cite{Poling:2006da},
and the underlying light-quark ``continuum'' is about 12 nb. Through
this paper, charge conjugate modes are implicitly assumed, unless
otherwise noted. 

\section{\boldmath Procedure}

\subsection{\boldmath $D^0$ and $D^+$}

Here we employ a single-tag (ST) technique extensively used by
CLEO-c~\cite{CLEO:sys, Peterpaper, FanKpi0, FanDspp}, pioneered by the
Mark~III Collaboration at SPEAR for measuring $D^0$ and $D^+$
branching fractions~\cite{mark3a, mark3b}, which exploits a feature of
near-threshold production of charmed mesons, {\it i.e.} 
$M_{\mathrm{bc}}$ and $\Delta E$, see below. 

We formed $D$ and $\bar{D}$ candidates in all $D \rightarrow PP$ decay
modes from combinations of $\pi^{\pm}$, $K^{\pm}$, $\pi^0$,
$K^{0}_{S}$, $\eta$, and $\eta'$ candidates selected using the
standardized requirements which are common to many CLEO-c analyses
involving $D$ decays. The $\psi$(3770) resonance is below the
kinematic threshold for $D \bar{D} \pi$ production, so the
events of interest, $e^{+} e^{-} \rightarrow \psi(3770) \rightarrow D
\bar{D}$, have $D$ mesons with energy equal to the beam
energy. Two variables reflecting energy and momentum conservation are
used to identify valid $D$ candidates. They are $\Delta E \equiv
\sum_{i} E_{i} - E_{\rm beam}$, and $M_{\mathrm {bc}}\equiv
\sqrt{E^{2}_{\rm beam} - (\sum_{i} \mathbf{p}_{i})^{2}}$, where
\noindent $E_{i},\ \mathbf{p}_{i}$ are the energy and momentum of the
decay products of a $D$ candidate. For a correct combination of
particles, $\Delta E$ will be consistent with zero, and the
beam-constrained mass $M_{\mathrm {bc}}$ will be consistent with the
$D$ mass. Candidates are rejected if they fail mode-dependent
$\Delta E$ requirements. If there is more than one candidate in a
particular $D$ or $\bar{D}$ decay mode, we choose the candidate with
the smallest $|\Delta E|$.

\subsection{\boldmath $D^+_s$}

Unlike $D\bar{D}$ threshold events, conventional $\Delta E$ and
$M_\text{bc}$ variables are no longer good variables for $D_s$ from
$D_s^{\ast +} D_s^-$ decays, as the $D_s$ can either be a primary or
secondary (from a $D^\ast_s$ decay), with different momentum. 
We use the reconstructed invariant mass of the $D_s$ candidate,
$M(D_s)$, and the mass recoiling against the $D_s$ candidate, 
$M_\text{recoil}(D_s)
\equiv \sqrt{ (E_{0} - E_{D_s} )^2 - (\mathbf{p}_{0}-\mathbf{p}_{D_s})^2 }
$, as our primary kinematic variables to select a $D_s$
candidate. Here $(E_{0},\mathbf{p}_{0})$ is the net four-momentum of
the $e^+e^-$ system, taking the finite beam crossing angle into account,
$\mathbf{p}_{D_s}$ is the momentum of the $D_s$ candidate, 
$E_{D_s} = \sqrt{m^2_{D_s} + \mathbf{p}^2_{D_s}}$,
and $m_{D_s}$ is the known $D_s$ mass~\cite{PDGValue}.
We make no requirements on the decay of the other $D_s$ in the event.

There are two components in the recoil mass distribution, a peak
around the $D^\ast_s$ mass if the candidate is due to the primary 
$D_s$ and a rectangular shaped distribution if the candidate is due to the
secondary $D_s$ from a $D^\ast_s$ decay. The edges of
$M_\text{recoil}(D_s)$ from the secondary $D_s$ are
kinematically determined (as a function of $\sqrt{s}$ and known masses),
and at $\sqrt{s} = 4170$~MeV, $\Delta M_\text{recoil}(D_s)
\equiv M_\text{recoil}(D_s) - m_{D^\ast_s}$ is in the range
$[-54, 57]$~MeV. Initial state radiation causes a tail on the high
side, above 57~MeV.
We select $D_s$ candidates within the 
$-55~\text{MeV} \le \Delta M_\text{recoil}(D_s) < +55~\text{MeV}$
range. This window allows both primary and secondary $D_s$
candidates to be selected.

We also require a photon consistent with coming from 
$D^{\ast +}_s \rightarrow D^+_s \gamma$ decay, by looking at the
mass recoiling against the $D_s$ candidate plus $\gamma$ system,
$M_\text{recoil}(D_s \gamma)
\equiv \sqrt{ (E_{0} - E_{D_s} - E_\gamma)^2 - (\mathbf{p}_{0}-\mathbf{p}_{D_s}-\mathbf{p}_\gamma)^2 }$.
For correct combinations, this recoil mass peaks at $m_{D_s}$, 
regardless of
whether the candidate is due to a primary or a secondary $D_s$.  
We require $| M_\text{recoil}(D_s \gamma) - m_{D_{s}} | <
30~\text{MeV}$. This requirement improves the
signal to noise ratio, important for the suppressed modes. Every event
is allowed to contribute a maximum of one $D_s$ candidate per mode and
charge. If there are multiple candidates, the one with
$M_\text{recoil}(D_s\gamma)$ closest to $m_{D_s}$ is chosen.

\subsection{\boldmath Common}

Our standard final-state particle selection requirements are described
in detail elsewhere~\cite{CLEO:sys}. Charged tracks produced in the
$D$ decay are required to satisfy criteria based on the track fit
quality, and angles $\theta$ with respect to the beam line,
satisfying $|\cos\theta|<0.93$. Momenta of charged particles utilized
in $D^0$ and $D^+$ candidate reconstructions must be above
50~MeV/$c$, while those for $D_s$ must be above 100~MeV/$c$ to
eliminate the soft pions from $D^\ast \bar{D}^\ast$ and $D^\ast \bar{D}$
decays (through $D^\ast \to \pi D$). Tracks must also be consistent
with their coming from the interaction point in three dimensions. Pion and
kaon candidates are required to have $dE/dx$ measurements within three
standard deviations ($3\sigma$) of the expected value. For tracks with
momenta greater than 700~MeV/$c$, RICH information, if available, is
combined with $dE/dx$.

The $K^0_S$ candidates are selected from pairs of oppositely-charged
and vertex-constrained tracks having invariant mass within
7.5~MeV, or roughly 3$\sigma$, of the known $K^0_S$ mass~\cite{PDGValue}.
We identify $\pi^{0}$ candidates via $\pi^{0} \rightarrow \gamma \gamma$,
detecting the photons in the CsI calorimeter. 
To avoid having both photons in a region of poorer energy resolution,
we require that at least one of the photons be in the ``good barrel''
region, $|\cos \theta_{\gamma}| < 0.80$. We require that a
calorimeter cluster has a measured energy above 30~MeV, has a
lateral distribution consistent with that from photons, and not be
matched to any charged track.
The invariant mass of the photon pair is
required to be within 3$\sigma$ ($\sigma\sim$ 6 MeV) of the
known $\pi^0$ mass. A $\pi^0$ mass constraint is imposed when $\pi^0$
candidates are used in further reconstruction.
We reconstruct $\eta$ candidates in the decay of 
$\eta \rightarrow \gamma \gamma$. Candidates are formed using a similar
procedure as for $\pi^0$ except that $\sigma\sim$ 12~MeV.
We reconstruct $\eta'$ candidates in the decay mode 
$\eta' \rightarrow \pi^+ \pi^- \eta $. We
require $|m_{\pi^+ \pi^- \eta} - m_{\eta'}|< 10$~MeV.

\section{\boldmath Results}

\begin{figure*}
  \includegraphics*[width=0.95\textwidth]{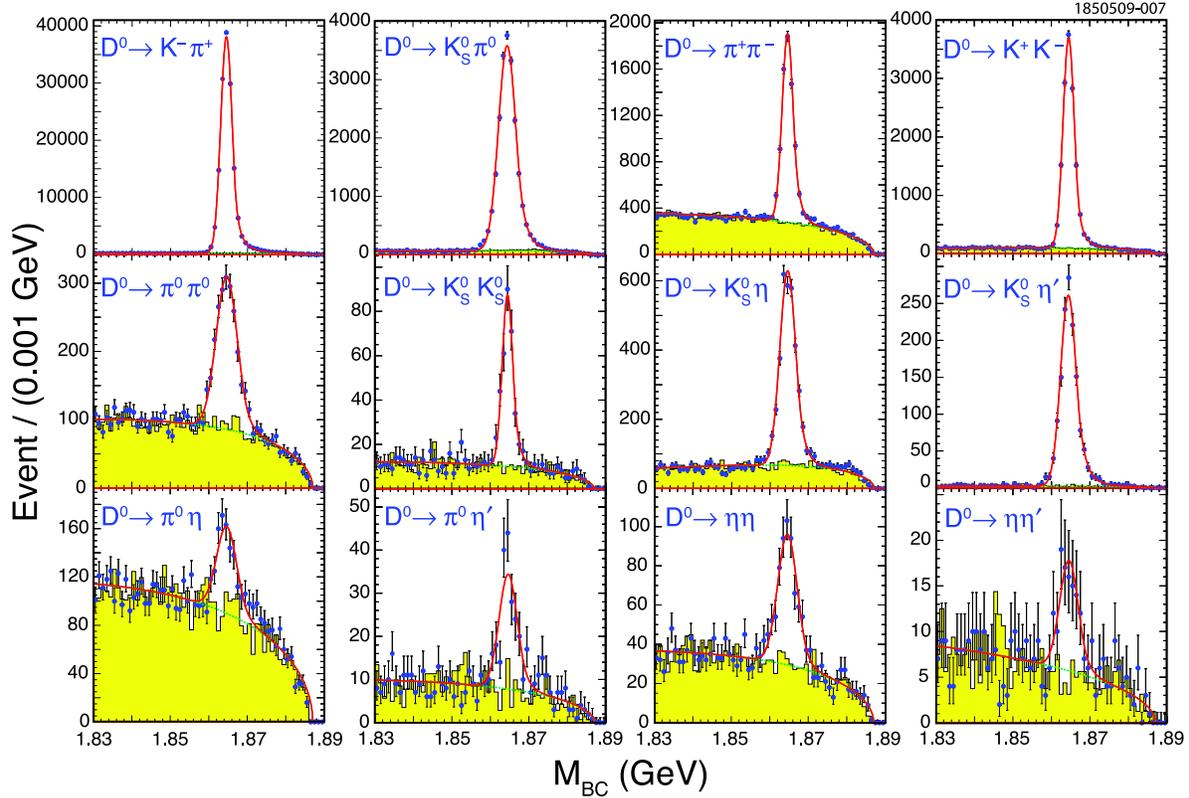}
  \caption{$M_{\mathrm {bc}}$ distributions of $D^0$ modes. For each
    distribution, the points are obtained from the $\Delta E$ signal
    region, the shaded histogram is from the $\Delta E$ sidebands, and
    the line is the fit.}
  \label{figure:one}
\end{figure*}

\begin{figure*}
  \includegraphics*[width=0.95\textwidth]{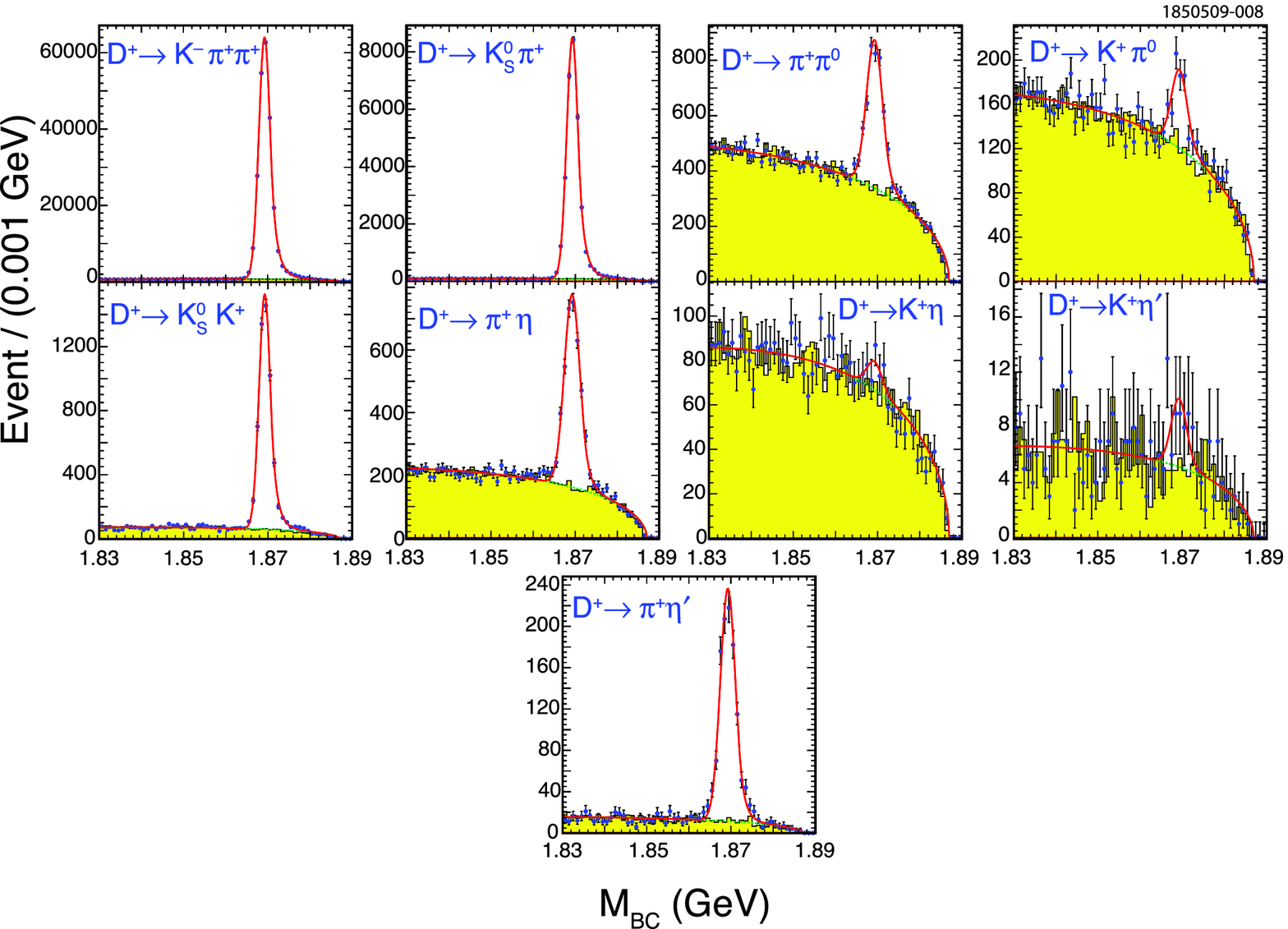}
  \caption{$M_{\mathrm {bc}}$ distributions of $D^+$ modes. For each
    distribution, the points are obtained from the $\Delta E$ signal
    region, the shaded histogram is from the $\Delta E$ sidebands, and
    the line is the fit.}
  \label{figure:two}
\end{figure*}

\begin{figure*}
  \includegraphics*[width=0.95\textwidth]{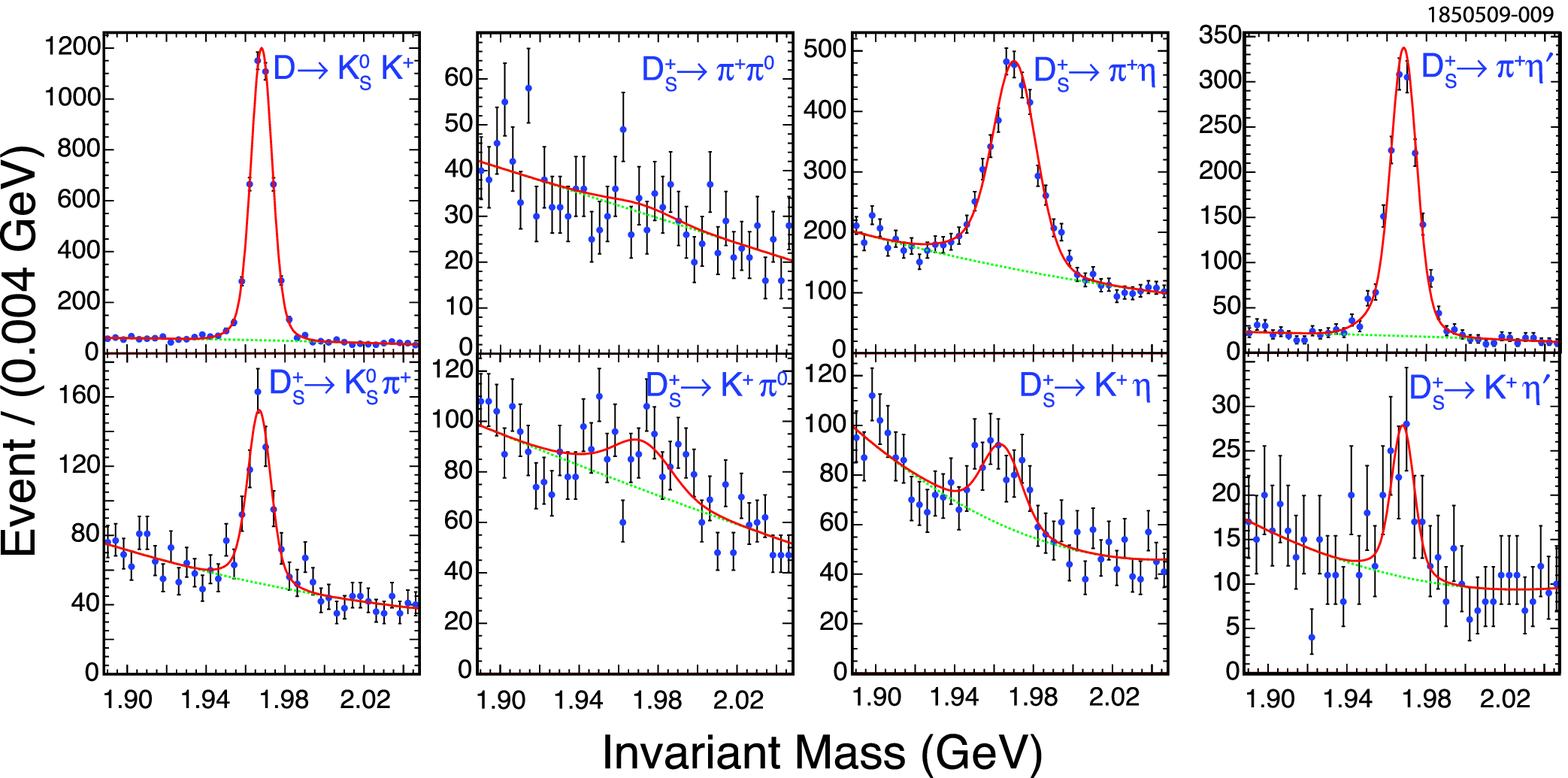}
  \caption{$M(D_s)$ distributions for $D_s$ modes. For each
    distribution, the points are the
    data and the superimposed line is the fit (the dotted line is the
    fitted background). The distribution for $D_s^+ \rightarrow \pi^+
    \pi^0$ has tighter requirements than the other modes -- see text.}
  \label{figure:three}
\end{figure*}

\subsection{\boldmath $D^0$ and $D^+$}

The $M_{\mathrm {bc}}$ distributions for the $D^0$ and $D^+$ candidate
combinations are shown in Figs.~\ref{figure:one} and~\ref{figure:two},
respectively. The points show the data and the lines are fits. The
normalization modes $D^0 \rightarrow K^- \pi^+$ and $D^+ \rightarrow
K^- \pi^+ \pi^+$ are essentially background-free. The backgrounds of
all modes are well described by the distributions obtained from the
$\Delta E$ sidebands. We perform a binned maximum likelihood fit to
extract the $D^0$ or $D^+$ signal yield from each $M_{\mathrm {bc}}$
distribution. For the signal, we use an inverted Crystal Ball line
shape \cite{CBFunc}, which is a Gaussian with a high-side tail. For
the background, we use an ARGUS function \cite{ArgusFunc}, with the shape
parameter determined from the $\Delta E$ sideband $M_{\mathrm {bc}}$
distribution, the high-end cutoff given by $E_{\rm{beam}}$, and the 
normalization determined from the fit to the $\Delta E$ signal
region. Results of the fits are shown in
Table~\ref{table:one}. Table~\ref{table:one} also includes the
detection efficiency for each
mode. The efficiencies include sub-mode branching
fractions~\cite{PDGValue} and have been corrected to include four
known small differences between data and Monte Carlo simulation, in
particular $\pi^0$-finding efficiency 0.96, $\eta$-finding efficiency
0.935, $\pi^{\pm}$ particle identification 0.995, and $K^{\pm}$
particle identification 0.99, data efficiency being smaller than MC
efficiency by those ratios.

\subsection{\boldmath $D^+_s$}

The resulting $M(D_s)$ distributions for $D_s$ modes are shown in
Fig.~\ref{figure:three}. The points show the data and the lines are
fits. We perform binned maximum likelihood fits to extract signal
yields from the $M(D_s)$ distributions. For the signal, we use the sum
of two Gaussians for the line shape. For the background, we use a
second-degree polynomial function. Results of the fits and detection
efficiencies are given in Table~\ref{table:one}.

\begin{table}[tb]
  \begin{center}
    \caption{\label{table:one}Observed yields from data and
      reconstruction efficiencies and their statistical
      uncertainties. The efficiencies include sub-mode branching
      fractions~\cite{PDGValue} and have been corrected to include
      several known small differences between data and Monte Carlo
      simulation.}
    \begin{tabular}{ l c c }  \hline \hline
      Mode  & Efficiency (\%)  & Yield\\ \hline

      $D^{0} \rightarrow K^{+} K^{-}$ & 57.35 $\pm$ 0.16 & 13782 $\pm$ 136 \\
      $D^{0} \rightarrow K^{0}_{S} K^{0}_{S}$ & 22.73 $\pm$ 0.13 & 215 $\pm$ 23 \\
      $D^{0} \rightarrow \pi^{+} \pi^{-}$ & 72.68 $\pm$ 0.14 & 6210 $\pm$ 93 \\
      $D^{0} \rightarrow \pi^{0} \pi^{0}$ & 32.95 $\pm$ 0.14 & 1567 $\pm$ 54 \\
      $D^{0} \rightarrow K^{-} \pi^{+}$ & 65.11 $\pm$ 0.15 & 150259 $\pm$ 420 \\
      $D^{0} \rightarrow K^{0}_{S} \pi^{0}$ & 28.57 $\pm$ 0.14 & 20045 $\pm$ 165 \\
      $D^{0} \rightarrow K^{0}_{S} \eta$ & 10.08 $\pm$ 0.05 & 2864 $\pm$ 65 \\
      $D^{0} \rightarrow \pi^{0} \eta$ & 11.97 $\pm$ 0.05 & 481 $\pm$ 40 \\
      $D^{0} \rightarrow K^{0}_{S} \eta'$ & 2.35 $\pm$ 0.02 & 1321 $\pm$ 42 \\
      $D^{0} \rightarrow \pi^{0} \eta'$ & 2.97 $\pm$ 0.02 & 159 $\pm$ 19 \\
      $D^{0} \rightarrow \eta \eta$ & 4.35 $\pm$ 0.02 & 430 $\pm$ 29 \\
      $D^{0} \rightarrow \eta \eta'$ & 1.06 $\pm$ 0.01 & 66 $\pm$ 15 \\ \hline

      $D^{+} \rightarrow K^{-} \pi^{+} \pi^{+}$ & 54.92 $\pm$ 0.16 & 231058 $\pm$ 515 \\
      $D^{+} \rightarrow K^{0}_{S} K^{+}$ & 36.62 $\pm$ 0.15 & 5161 $\pm$ 86 \\
      $D^{+} \rightarrow \pi^{+} \pi^{0}$ & 48.69 $\pm$ 0.15 & 2649 $\pm$ 76 \\
      $D^{+} \rightarrow K^{0}_{S} \pi^{+}$ & 42.54 $\pm$ 0.16 & 30095 $\pm$ 191 \\
      $D^{+} \rightarrow K^{+} \pi^{0}$ & 43.29 $\pm$ 0.15 & 343 $\pm$ 37 \\
      $D^{+} \rightarrow K^{+} \eta$ & 15.95 $\pm$ 0.06 & 60 $\pm$ 24 \\
      $D^{+} \rightarrow \pi^{+} \eta$ & 18.07 $\pm$ 0.06 & 2940 $\pm$ 68 \\
      $D^{+} \rightarrow K^{+} \eta'$ & 4.29 $\pm$ 0.02 & 23 $\pm$ 18 \\
      $D^{+} \rightarrow \pi^{+} \eta'$ & 4.81 $\pm$ 0.02 & 1037 $\pm$ 35 \\ \hline

      $D_{s}^{+} \rightarrow K^{0}_{S} K^{+}$ & 24.73 $\pm$ 0.14 & 4076 $\pm$ 71 \\
      $D_{s}^{+} \rightarrow \pi^{+} \pi^{0}$ & 16.60 $\pm$ 0.12 & 19 $\pm$ 28 \\
      $D_{s}^{+} \rightarrow K^{0}_{S} \pi^{+}$ & 28.15 $\pm$ 0.14 & 393 $\pm$ 33 \\
      $D_{s}^{+} \rightarrow K^{+} \pi^{0}$ & 29.57 $\pm$ 0.14 & 202 $\pm$ 70 \\
      $D_{s}^{+} \rightarrow K^{+} \eta$ & 11.40 $\pm$ 0.05 & 222 $\pm$ 41 \\
      $D_{s}^{+} \rightarrow \pi^{+} \eta$ & 12.70 $\pm$ 0.06 & 2587 $\pm$ 89 \\
      $D_{s}^{+} \rightarrow K^{+} \eta'$ & 2.87 $\pm$ 0.02 & 56 $\pm$ 17 \\
      $D_{s}^{+} \rightarrow \pi^{+} \eta'$ & 3.28 $\pm$ 0.02 & 1436 $\pm$ 47 \\

      \hline \hline
    \end{tabular}
  \end{center}
\end{table}      

\subsection{\boldmath Upper Limits}

For most of the $D \rightarrow PP$ modes, very clear signals are found
in data. We find no significant evidence for $D^{+} \rightarrow K^{+}
\eta$, $D^{+} \rightarrow K^{+} \eta'$, and $D_{s}^{+} \rightarrow
\pi^{+} \pi^{0}$ decays, and therefore set upper limits on their
branching fractions. The $M_{\mathrm {bc}}$ distributions of $D^{+}
\rightarrow K^{+} \eta$ and $D^{+} \rightarrow K^{+} \eta'$ modes are
shown in Fig.~\ref{figure:two}.
Monte Carlo studies indicate that tightening the requirements on
$M_\text{recoil}(D_s)$ to $\pm 10$~MeV and $M_\text{recoil}(D_s
\gamma)$ to $\pm 20$~MeV should improve the upper limit on $D^+_s
\rightarrow \pi^+ \pi^0$ decay. Consequently, for $D^+_s \rightarrow
\pi^+ \pi^0$ (and only $D^+_s \rightarrow \pi^+ \pi^0$), we have
applied these tighter requirements. The invariant mass distribution
for $D^+_s \rightarrow \pi^+ \pi^0$ shown in Fig.~\ref{figure:three}
and the efficiency given in Table~\ref{table:one} have these tighter
requirements.

\subsection{\boldmath Background from Non-resonant Decays}

Non-resonant $D$ decays can enter into our signal
modes with the same final particles. For example, non-resonant 
$D^+ \rightarrow \pi^+ (\pi^+ \pi^-)$ can appear in the 
$D^+ \rightarrow \pi^+ K^{0}_{S}, K^{0}_{S} \rightarrow \pi^+ \pi^-$
mode. Also, non-resonant $D^+ \rightarrow \pi^+ (\pi^+ \pi^- \eta)$
can appear in the $D^+ \rightarrow \pi^+ \eta', \eta' \rightarrow
\pi^+ \pi^- \eta$ mode. 
To understand the backgrounds from non-resonant $D^0$ or $D^+$ decays,
we look at $M_{\mathrm {bc}}$ distributions in the invariant mass
sideband regions of the intermediate resonances ($K^{0}_{S}$ or
$\eta'$). For $D^+_s$ decays, we follow the same procedure, 
replacing $M_{\mathrm {bc}}$ with $M(D_s)$. The scaling factor, from
sideband to signal region, is taken to be unity, as indicated by Monte
Carlo studies.

For the $D^0 \rightarrow K^{0}_{S} K^{0}_{S}$ (or $D^0
\rightarrow K^{0}_{S} \eta'$) mode, the scatter plot of $K^{0}_{S}$
candidate invariant mass against the other $K^{0}_{S}$ (or $\eta'$)
candidate invariant mass is used to define a signal region and two
kinds of sideband regions to remove the non-resonant decay
background. Again, the scaling factor, from sideband to signal region,
is taken to be unity.

\subsection{\boldmath Systematic Uncertainties}

We have considered several sources of systematic uncertainty. Some are
correlated among different decay modes. These include: 
\begin{enumerate}
\item the uncertainty associated with the efficiency for finding a
track - 0.3\% per track~\cite{CLEO:sys};
\item an additional 0.6\% per kaon track is added~\cite{CLEO:sys},
  uncorrelated with item 1;
\item the uncertainty in charged pion identification is 0.3\% per
  $\pi^{\pm}$~\cite{CLEO:sys};
\item the uncertainty in charged kaon identification is 0.3\% per
  $K^{\pm}$~\cite{CLEO:sys}, uncorrelated with item 3;
\item the relative systematic uncertainties for $\pi^0$, $K^0_S$, and
  $\eta$ finding efficiencies are 2.0\%, 1.8\%~\cite{CLEO:sys}, and
  4.0\%, independent of one another, and independent of the first
  four-mentioned uncertainties;
\item finally, among the correlated systematic uncertainties, there
  are the uncertainties in the input branching fractions of the
  normalization modes, 2.0\% for $D^{0} \rightarrow K^{-}
  \pi^{+}$~\cite{CLEO:sys}, 2.2\% for $D^{+} \rightarrow K^{-} \pi^{+}
  \pi^{+}$~\cite{CLEO:sys}, and 5.8\% for $D_{s}^{+} \rightarrow
  K^{0}_{S} K^{+}$~\cite{Peterpaper}.
\end{enumerate}
Note that for $K^0_S$, with $K^0_S \rightarrow \pi^+ \pi^-$, item 1
applies, as the tracks must be found, but item 3 does not apply, as
pion identification is not required for $K^0_S \rightarrow \pi^+
\pi^-$.

The systematic uncertainties that are uncorrelated among the decay
modes include those due to choice of signal shape and background
shape. They range from $\pm 0.05\%$ for the cleaner decay modes to
$\pm 4.55\%$ for the modes with substantial background.

In the Table~\ref{table:two} we separately list, for each decay mode,
the quadratic sum of the systematic errors excluding that from the
normalization mode, and the error from the uncertainty in the
normalization mode.

\subsection{\boldmath $CP$ Asymmetries}

The Standard Model predicts that direct $CP$ violation in $D$ decays,
{\it e.g.}, a difference in the branching fractions for 
$D^{+}_{s} \rightarrow K^+ \eta$ and $D^{-}_{s} \rightarrow K^- \eta$,
will be vanishingly small. We have separate yields and efficiencies
for $D$ and $\bar{D}$ events, so it is possible to compute asymmetries
$\mathcal{A}_{CP} \equiv (\mathcal{B}_{+} - \mathcal{B}_{-}) /
(\mathcal{B}_{+} + \mathcal{B}_{-})$, which are sensitive to direct
$CP$ violation in $D$ decays. All systematic uncertainties cancel in
this ratio, with the exception of charged pion and kaon tracking and
particle identification efficiencies. Here the relative factor is the charge
dependence of the efficiencies in data and Monte Carlo
simulations~\cite{CLEO:sys}.

For $D^0$ vs.\ $\bar{D^0}$, the only asymmetry we can measure is
$K^-\pi^+$ {\it vs.} $K^+\pi^-$. That difference will contain a component
from the difference in the doubly-Cabibbo-suppressed decays $D^0
\rightarrow K^+ \pi^-$ {\it vs.} $\bar{D^0} \rightarrow K^- \pi^+$, as well
as the component from the favored decays $D^0 \rightarrow K^- \pi^+$
{\it vs.} $\bar{D^0} \rightarrow K^+ \pi^-$. Our measurement does not
separate these two possible asymmetries.

\section{\boldmath Summary}

The obtained branching ratios, branching fractions, and $CP$
asymmetries for all $D \rightarrow PP$ modes are shown in
Table~\ref{table:two}. The values
we obtained are consistent with the world averages~\cite{PDGValue} and
for the suppressed modes, of better accuracy. No significant $CP$
asymmetries are observed.

\begin{table*}[tb]
  \begin{center}
    \caption{\label{table:two}Ratios of branching fractions to the
      corresponding normalization modes $D^{0} \rightarrow K^{-}
      \pi^{+}$, $D^{+} \rightarrow K^{-} \pi^{+} \pi^{+}$, and
      $D_{s}^{+} \rightarrow K^{0}_{S} K^{+}$; branching fractions
      results from this analysis; and charge asymmetries
      $\mathcal{A}_{CP}$. Uncertainties are statistical error,
      systematic error, and the error from the input branching
      fractions of normalization modes. (For $D^0$, the normalization
      mode is the sum of $D^0 \rightarrow K^- \pi^+$ and $D^0
      \rightarrow K^+ \pi^-$ -- see text.)}
    \begin{tabular}{ l c c c}  \hline \hline
      Mode
      & $\mathcal{B}_{\text{mode}}/\mathcal{B}_{\text{Normalization}}$ (\%)
      & This result $\mathcal{B}$ (\%)
      & $\mathcal{A}_{CP}$ (\%) \\ \hline

      $D^{0} \rightarrow K^{+} K^{-}$ & 10.41 $\pm$ 0.11 $\pm$ 0.11 & 0.407 $\pm$ 0.004 $\pm$ 0.004 $\pm$ 0.008 &  \\
      $D^{0} \rightarrow K^{0}_{S} K^{0}_{S}$ & 0.41 $\pm$ 0.04 $\pm$ 0.02 & 0.0160 $\pm$ 0.0017 $\pm$ 0.0008 $\pm$ 0.0003 &  \\
      $D^{0} \rightarrow \pi^{+} \pi^{-}$ & 3.70 $\pm$ 0.06 $\pm$ 0.09 & 0.145 $\pm$ 0.002 $\pm$ 0.004 $\pm$ 0.003 &  \\
      $D^{0} \rightarrow \pi^{0} \pi^{0}$ & 2.06 $\pm$ 0.07 $\pm$ 0.10 & 0.081 $\pm$ 0.003 $\pm$ 0.004 $\pm$ 0.002 &  \\
      $D^{0} \rightarrow K^{-} \pi^{+}$ & 100 & 3.9058 external input~\cite{CLEO:sys} & 0.5 $\pm$ 0.4 $\pm$ 0.9 \\
      $D^{0} \rightarrow K^{0}_{S} \pi^{0}$ & 30.4 $\pm$ 0.3 $\pm$ 0.9 & 1.19 $\pm$ 0.01 $\pm$ 0.04 $\pm$ 0.02 &  \\
      $D^{0} \rightarrow K^{0}_{S} \eta$ & 12.3 $\pm$ 0.3 $\pm$ 0.7 & 0.481 $\pm$ 0.011 $\pm$ 0.026 $\pm$ 0.010 &  \\
      $D^{0} \rightarrow \pi^{0} \eta$ & 1.74 $\pm$ 0.15 $\pm$ 0.11 & 0.068 $\pm$ 0.006 $\pm$ 0.004 $\pm$ 0.001 &  \\
      $D^{0} \rightarrow K^{0}_{S} \eta'$ & 24.3 $\pm$ 0.8 $\pm$ 1.1 & 0.95 $\pm$ 0.03 $\pm$ 0.04 $\pm$ 0.02 &  \\
      $D^{0} \rightarrow \pi^{0} \eta'$ & 2.3 $\pm$ 0.3 $\pm$ 0.2 & 0.091 $\pm$ 0.011 $\pm$ 0.006 $\pm$ 0.002 &  \\
      $D^{0} \rightarrow \eta \eta$ & 4.3 $\pm$ 0.3 $\pm$ 0.4 & 0.167 $\pm$ 0.011 $\pm$ 0.014 $\pm$ 0.003 &  \\
      $D^{0} \rightarrow \eta \eta'$ & 2.7 $\pm$ 0.6 $\pm$ 0.3 & 0.105 $\pm$ 0.024 $\pm$ 0.010 $\pm$ 0.002 &  \\ \hline

      $D^{+} \rightarrow K^{-} \pi^{+} \pi^{+}$ & 100 & 9.1400 external input~\cite{CLEO:sys} & -0.1 $\pm$ 0.4 $\pm$ 0.9 \\
      $D^{+} \rightarrow K^{0}_{S} K^{+}$ & 3.35 $\pm$ 0.06 $\pm$ 0.07 & 0.306 $\pm$ 0.005 $\pm$ 0.007 $\pm$ 0.007 & -0.2 $\pm$ 1.5 $\pm$ 0.9 \\
      $D^{+} \rightarrow \pi^{+} \pi^{0}$ & 1.29 $\pm$ 0.04 $\pm$ 0.05 & 0.118 $\pm$ 0.003 $\pm$ 0.005 $\pm$ 0.003 & 2.9 $\pm$ 2.9 $\pm$ 0.3 \\
      $D^{+} \rightarrow K^{0}_{S} \pi^{+}$ & 16.82 $\pm$ 0.12 $\pm$ 0.37 & 1.537 $\pm$ 0.011 $\pm$ 0.034 $\pm$ 0.033 & -1.3 $\pm$ 0.7 $\pm$ 0.3 \\
      $D^{+} \rightarrow K^{+} \pi^{0}$ & 0.19 $\pm$ 0.02 $\pm$ 0.01 & 0.0172 $\pm$ 0.0018 $\pm$ 0.0006 $\pm$ 0.0004 & -3.5 $\pm$ 10.7 $\pm$ 0.9 \\
      $D^{+} \rightarrow K^{+} \eta$ & $<0.14$ (90\% C.L.) & $<0.013$ (90\% C.L.) &  \\
      $D^{+} \rightarrow \pi^{+} \eta$ & 3.87 $\pm$ 0.09 $\pm$ 0.19 & 0.354 $\pm$ 0.008 $\pm$ 0.018 $\pm$ 0.008 & -2.0 $\pm$ 2.3 $\pm$ 0.3 \\
      $D^{+} \rightarrow K^{+} \eta'$ & $<0.20$ (90\% C.L.) & $<0.018$ (90\% C.L.) &  \\
      $D^{+} \rightarrow \pi^{+} \eta'$ & 5.12 $\pm$ 0.17 $\pm$ 0.25 & 0.468 $\pm$ 0.016 $\pm$ 0.023 $\pm$ 0.010 & -4.0 $\pm$ 3.4 $\pm$ 0.3 \\ \hline

      $D_{s}^{+} \rightarrow K^{0}_{S} K^{+}$ & 100 & 1.4900 external input~\cite{Peterpaper} & 4.7 $\pm$ 1.8 $\pm$ 0.9 \\
      $D_{s}^{+} \rightarrow \pi^{+} \pi^{0}$ & $<2.3$ (90\% C.L.) & $<0.037$ (90\% C.L.) &  \\
      $D_{s}^{+} \rightarrow K^{0}_{S} \pi^{+}$ & 8.5 $\pm$ 0.7 $\pm$ 0.2 & 0.126 $\pm$ 0.011 $\pm$ 0.003 $\pm$ 0.007 & 16.3 $\pm$ 7.3 $\pm$ 0.3 \\
      $D_{s}^{+} \rightarrow K^{+} \pi^{0}$ & 4.2 $\pm$ 1.4 $\pm$ 0.2 & 0.062 $\pm$ 0.022 $\pm$ 0.004 $\pm$ 0.004 & -26.6 $\pm$ 23.8 $\pm$ 0.9 \\
      $D_{s}^{+} \rightarrow K^{+} \eta$ & 11.8 $\pm$ 2.2 $\pm$ 0.6 & 0.176 $\pm$ 0.033 $\pm$ 0.009 $\pm$ 0.010 & 9.3 $\pm$ 15.2 $\pm$ 0.9 \\
      $D_{s}^{+} \rightarrow \pi^{+} \eta$ & 123.6 $\pm$ 4.3 $\pm$ 6.2 & 1.84 $\pm$ 0.06 $\pm$ 0.09 $\pm$ 0.11 & -4.6 $\pm$ 2.9 $\pm$ 0.3 \\
      $D_{s}^{+} \rightarrow K^{+} \eta'$ & 11.8 $\pm$ 3.6 $\pm$ 0.6 & 0.18 $\pm$ 0.05 $\pm$ 0.01 $\pm$ 0.01 & 6.0 $\pm$ 18.9 $\pm$ 0.9 \\
      $D_{s}^{+} \rightarrow \pi^{+} \eta'$ & 265.4 $\pm$ 8.8 $\pm$ 13.9 & 3.95 $\pm$ 0.13 $\pm$ 0.21 $\pm$ 0.23 & -6.1 $\pm$ 3.0 $\pm$ 0.3 \\

     \hline \hline
    \end{tabular}
  \end{center}
\end{table*}      

\section{\boldmath Acknowledgements}

We gratefully acknowledge the effort of the CESR staff
in providing us with excellent luminosity and running conditions.
D.~Cronin-Hennessy and A.~Ryd thank the A.P.~Sloan Foundation.
This work was supported by the National Science Foundation,
the U.S. Department of Energy,
the Natural Sciences and Engineering Research Council of Canada, and
the U.K. Science and Technology Facilities Council.

\end{document}